\documentclass[aps, prb, twocolumn,amsmath,amssymb,floatfix, reprint, longbibliography,superscriptaddress]{revtex4-2}

\usepackage{amsfonts,amsmath,amssymb,amstext,dsfont,bm}
\usepackage{graphicx}
\usepackage{svg}
\usepackage{times}
\usepackage{dcolumn}
\usepackage{hyperref}
\usepackage[normalem]{ulem}
\usepackage{chemformula}
\hypersetup{colorlinks, allcolors=blue}

\newcommand{\sign}[1]{\,\mbox{sgn}\left({#1}\right)}

\newcommand{\RE}[1]{\,\mbox{Re}{\,#1}}
\newcommand{\IM}[1]{\,\mbox{Im}{\,#1}}
\newcommand{\df}[1]{\,\delta{\left(#1\right)}}
\definecolor{Red}{rgb}{1,0,0}
\definecolor{Blue}{rgb}{0.5,0,0.9}
\definecolor{Green}{rgb}{0,0.5,0}
\definecolor{Magenta}{rgb}{1,0,0.56}
\definecolor{Orange}{rgb}{1,0.64,0}

\begin{document}
\title{Effects of the Hubbard interaction on the quantum metric}

\author{Pavlo Sukhachov}
\email{pavlo.sukhachov@ntnu.no}
\thanks{These authors contributed equally to this work}
\affiliation{Center for Quantum Spintronics, Department of Physics, Norwegian \\ University of Science and Technology, NO-7491 Trondheim, Norway}

\author{Niels Henrik Aase}
\email{niels.h.aase@ntnu.no}
\thanks{These authors contributed equally to this work}
\affiliation{Center for Quantum Spintronics, Department of Physics, Norwegian \\ University of Science and Technology, NO-7491 Trondheim, Norway}

\author{Kristian M{\ae}land}
\email{kristian.maeland@uni-wuerzburg.de}
\affiliation{Center for Quantum Spintronics, Department of Physics, Norwegian \\ University of Science and Technology, NO-7491 Trondheim, Norway}
\affiliation{Institute for Theoretical Physics and Astrophysics, University of W{\"u}rzburg, D-97074 W{\"u}rzburg, Germany}

\author{Asle Sudb{\o}}
\email{asle.sudbo@ntnu.no}
\affiliation{Center for Quantum Spintronics, Department of Physics, Norwegian \\ University of Science and Technology, NO-7491 Trondheim, Norway}

\date{February 24, 2025}

\begin{abstract}
Quantum geometry provides important information about the structure and topology of quantum states in various forms of quantum matter. The information contained therein has profound effects on observable quantities such as superconducting weight, Drude weight, and optical responses. Motivated by the recent advances in flat-band interacting systems, we investigate the role of interaction effects on the quantum metric. By using the fermionic Creutz ladder as a representative system, we show that the repulsive Hubbard interaction monotonically suppresses the quantum metric. While the eigenstates and their overlap quantifying the quantum metric can be obtained exactly in the presence of interactions through exact diagonalization, this method is limited to small system sizes. Alternatively, two theoretical proposals, the generalized quantum metric and the dressed quantum metric, suggest using renormalized Green's functions to define the interacting quantum metric. By comparing these analytical approaches with results from exact diagonalization, we show that the dressed quantum metric provides a better fit to the exact diagonalization results. Our conclusion holds for both flat-band and dispersive systems.
\end{abstract}

\maketitle
\section{Introduction}

Understanding the connection between the structure of the eigenstates of any physical system, namely, their topology, and various physical responses has recently become one of the key themes of modern condensed-matter physics. As in the general theory of relativity, the key notion in quantifying the topology is the quantum distance between the states, which, in general, is a complex tensor known as the quantum geometric tensor (QGT)~\cite{Provost1980a, Cheng:2010-rev, Rossi:2021-rev}.

The imaginary part of the QGT is the antisymmetric Berry curvature tensor~\cite{Berry:1984} related to an emergent gauge field, i.e., the Berry connection. The Berry curvature is manifested in several observable effects including the quantum Hall effect~\cite{Thouless-DenNijs-QuantizedHallConductance-1982}, anomalous Hall effect~\cite{Xiao-Niu:2009, Nagaosa:2010}, quantum oscillations~\cite{Mikitik-Sharlai-MagneticSusceptibilityTopological-2019}, etc. Systems where such effects can be observed include topological insulators, Dirac and Weyl semimetals, topological superconductors, optical lattices, cold atoms, magnons, etc.; see numerous books and reviews, e.g., Refs.~\cite{Hasan-Kane:rev-2010, Franz:book-2013, Bernevig2013, Cayssol2021, Armitage:rev-2018, Zhang-Zhu-TopologicalQuantumMatter-2018, GMSS:book, McClarty-TopologicalMagnonsReview-2022}.

The real part of the QGT is a symmetric tensor that defines the quantum metric and the distances between states. Unlike the Berry curvature, the role of the quantum metric in condensed-matter physics remains less studied. It was recently understood that the quantum metric plays an important role in several physical effects including, e.g., superconductivity and superfluidity~\cite{Peotta-Torma-SuperfluidityTopologicallyNontrivial-2015, Liang2017a, Rossi:2021-rev, Torma-Bernevig:2022-rev, Tian-Bockrath-EvidenceDiracFlat-2023, Shavit-Alicea-QuantumGeometricUnconventional-2024, Jahin-Lin-EnhancedKohnLuttingerTopological-2024}, optical responses~\cite{Gao-Xiao:2019-metrics, Ahn-Nagaosa:2020}, collective modes~\cite{Arora-Song:2022}, etc. The rise of interest in the quantum metric is also motivated by recent advances in creating narrow-band and flat-band systems exemplified by twisted bilayer graphene, see Ref.~\cite{Andrei-MacDonald:2020-rev} for a review. Other systems such as qubits and kagome metals could also demonstrate topological flat bands. The full quantum geometric tensor in these systems was recently measured in Refs.~\cite{Tan-Yu-ExperimentalMeasurementQuantum-2019, Yu-Cai-ExperimentalMeasurementQuantum-2020, Gianfrate-Malpuech-MeasurementQuantumGeometric-2020, Kang-Comin-MeasurementsQuantumGeometric-2024}.

Flat bands are particularly attractive for addressing the effects of interactions since, by definition, the potential energy dominates the kinetic one. Without interactions, even topological flat bands are less interesting since the corresponding quasiparticles are localized and do not support direct current electric conductivity~\cite{Huhtinen-Torma-ConductivityFlatBands-2023}. However, even in systems where interactions are pivotal for emergent physics, the quantum geometry of noninteracting states is typically used. In principle, the extension to interacting systems is straightforward by replacing single-particle states with many-body ground states~\cite{Souza-Martin:2000, Ozawa2019Nov, Salerno-Torma:2023} in the definition of the quantum metric. Approximate methods used to compute many-body ground states in the context of the quantum metric include Bogolyubov wave functions~\cite{Matsyshyn-Song-SuperconductingBerryCurvature-2024} and Hartree-Fock renormalized Bloch wave functions~\cite{Mao-Chowdhury-LowenergyOpticalAbsorption-2024}. Since many-body ground states are often challenging to obtain exactly, other many-body quantities were related to the single-particle quantum metric instead. For example, it was suggested that there is a relation between the many-body gap and the difference between the trace of the quantum metric and the magnitude of the Berry curvature, which quantifies the stability of fractional Chern insulators~\cite{Roy2014, Jackson2015, Andrews-Roy-StabilityFractionalChern-2024}. Still, the direct connection of these quantities to the interacting quantum metric is unclear.

The authors of Refs.~\cite{Chen-vonGersdorff:2022, Kashihara2023Mar, Onishi-Fu-QuantumWeight-2024} proposed alternative extensions of the quantum metric to interacting systems based on Green's functions. Indeed, single-particle Green's functions were already used before to define topological invariants, such as the Thouless-Kohmoto-Nightingale-den Nijs invariant~\cite{Volovik:book-2003}. In Ref.~\cite{Chen-vonGersdorff:2022}, the authors introduced the dressed quantum metric (DQM) defined via the charge polarization susceptibility linked to the dynamic fidelity susceptibility~\cite{You-Gu-FidelityDynamicStructure-2007, Gu-FidelityApproachQuantum-2010}. Interaction effects are included via self-energy corrections in the corresponding Green's functions and could be probed via the spectral function. On the other hand, Refs.~\cite{Kashihara2023Mar, Onishi-Fu-QuantumWeight-2024} use a slightly different approach where the quantum metric is defined via a current-current correlation function~\cite{Souza-Martin:2000} and is directly related to the optical conductivity. Again, the effects of interactions are included via self-energy corrections in the Green's functions. The corresponding quantum metric is dubbed the generalized quantum metric (GQM)~\cite{Kashihara2023Mar}.

Both DQM~\cite{Chen-vonGersdorff:2022} and GQM~\cite{Kashihara2023Mar} provide the same quantum metric in the noninteracting limit. While this is a crucial check any definition of the quantum metric should satisfy, it is unclear to which extent these approaches agree in the presence of interactions. Furthermore, to our knowledge no comparison with a direct calculation of the quantum metric in the presence of interactions, using, e.g., exact diagonalization, has been performed.

In our work, we analytically and numerically analyze the quantum metric in interacting systems exemplified by a fermionic Creutz ladder~\cite{Creutz1999}. We emphasize that our work differs from Ref.~\cite{Salerno-Torma:2023} where a corresponding bosonic system was studied numerically. The Creutz ladder is a paradigmatic one-dimensional (1D) model allowing for only two flat bands; other models either have both dispersive and flat bands (e.g., a sawtooth lattice) or have more than two bands (e.g., stub, diamond, and cross lattices). The model can be simulated via ultracold atoms in optical lattices~\cite{Mazza-Lewenstein-OpticallatticebasedQuantumSimulator-2012, Junemann2017, Kang2020ExpCreutz}. In the analytically accessible flat-band limit, we explicitly show that the DQM and the GQM differ in interacting systems. While both approaches deviate from the numerical results obtained via exact diagonalization, the DQM provides a much better fit to the exact diagonalization results compared with the GQM advocated in Ref.~\cite{Kashihara2023Mar}. This result persists for flat and weakly dispersive bands suggesting that the DQM provides a more viable definition of the many-body quantum metric. In addition to these findings, we investigate the manifestation of the interaction effects in the spectral function and the optical conductivity. Interactions lead to shifts of the spectral peaks present in the noninteracting system and the appearance of additional peaks in the spectral function away from the bare bands, known as replica bands \cite{Schrodi2021Replica, Chen2024}.

This paper is organized as follows. In Sec.~\ref{sec:creutz}, we introduce the fermionic Creutz ladder with Hubbard interactions, define Green's functions, and calculate self-energy corrections. In Sec.~\ref{sec:quantum_metric}, we use this to derive the expressions for the proposed interacting quantum metrics, which we then benchmark against results obtained via exact diagonalization. In Sec.~\ref{sec:results}, we present our results, and  also discuss the manifestations of the interaction effects in the spectral properties of the system. We summarize our work in Sec.~\ref{sec:conclusion}. Technical aspects of calculations are presented in Appendixes~\ref{app:self-energy} and \ref{app:g_steps}. In this work, we use natural units $\hbar=c=k_\mathrm{B}=1$ and take the lattice constant to be unity $a=1$.

\section{Fermionic Creutz ladder with Hubbard interactions}
\label{sec:creutz}

\subsection{Model}
\label{sec:creutz-model}

\begin{figure}
    \centering
    \includegraphics[width=\linewidth]{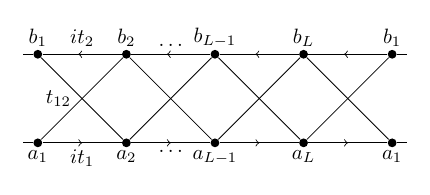}
    \caption{Illustration of the Creutz ladder with periodic boundary conditions. Hopping along the diagonals is given by $t_{12}$, and there is no hopping across the ladder rungs. The hopping constant along the ladder legs is purely imaginary, $\pm it_{1}$ and $\pm it_{2}$, with the arrows showing its sign in the case of spin up, and $t_1$ ($t_2$) being the hopping strength between the lower (upper) sites.}
    \label{fig:creutz_ladder_illustration}
\end{figure}

To present our results in a simple and concise form, we employ a one-dimensional (1D) model known as the Creutz ladder~\cite{Creutz1999}. It is a tunable model that can realize flat and dispersive bands. Specifically, we consider a spinful fermionic Creutz ladder \cite{Tovmasyan2016Dec}, with Hamiltonian
\begin{align}
    \hat{H} &= \frac{1}{2}\sum_{j\sigma} \bigg(i\sigma t_1 a_{j+1, \sigma}^{\dag}a_{j,\sigma} -i\sigma t_2 b_{j+1,\sigma}^{\dag}b_{j,\sigma} \nonumber \\
    & +t_{12}a_{j+1,\sigma}^{\dag}b_{j,\sigma} + t_{12}b_{j+1,\sigma}^{\dag}a_{j,\sigma} +\text{H.c.}\bigg).
\label{model_H}
\end{align}
Here, $t_1, t_{2}$, and $t_{12}$ are the hopping strengths along the lower ladder leg, upper ladder leg, and along the diagonals, respectively. Furthermore, $a_{j,\sigma}$ and $b_{j,\sigma}$ ($a^\dagger_{j,\sigma}$ and $b^\dagger_{j,\sigma}$) are the annihilation (creation) operators of an electron located at lattice site $j$ with spin $\sigma$ on the lower and upper legs of the ladder, respectively; see Fig.~\ref{fig:creutz_ladder_illustration} for an illustration.~\footnote{Note that the sawtooth lattice~\cite{Nakamura-Kubo-ElementaryExcitationsDelta-1996, Sen-Cava-QuantumSolitonsSawtooth-1996} has a similar structure albeit different hopping constants.} The hopping along the ladder legs is taken to be complex. Physically, such a complex hopping can be induced by applying a magnetic field perpendicular to the ladder and choosing a gauge where the flux is picked up by moving along the upper and lower horizontal bonds~\cite{Creutz1999}. The length of the ladder is $L$ and the total number of lattice sites is $N=2L$.

\begin{figure}
    \centering
    \includegraphics[width=\linewidth]{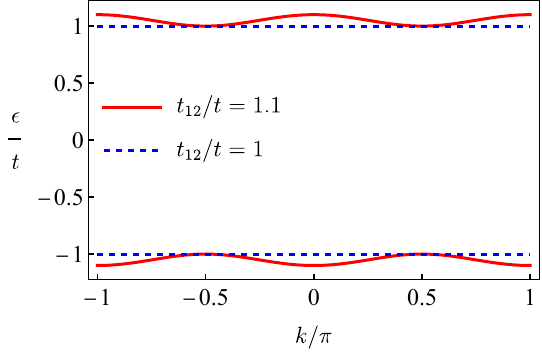}
    \caption{The energy spectrum (\ref{disp_general}) for two values of $t_{12}$ at $t_1=t_2 = t$ and $\sigma=+$.
    }
    \label{fig:creutz_ladder-spectrum}
\end{figure}

Introducing Fourier-transformed operators ${a_{k,\sigma}\equiv 1/\sqrt{L}\sum_j \mathrm{e}^{ikr_j} a_{j,\sigma}}$, as well as the vector $\boldsymbol{d}_k = (a_{k,\uparrow}, b_{k,\uparrow}, a_{k,\downarrow}, b_{k,\downarrow})^T$, we express $\hat{H}$ in a concise manner
\begin{equation}
    \hat{H} = \sum_k \boldsymbol{d}_k^\dagger H(k) \boldsymbol{d}_k \label{H_orig},
\end{equation}
where we introduced
\begin{align}
    H(k) &= \begin{pmatrix}
        H_{+}(k) & 0 \\
        0 & H_{-}(k)
    \end{pmatrix},
    \label{H-k-spin}\\
    H_\sigma(k) &\equiv
    \begin{pmatrix}
    -\sigma t_1  \sin k &  t_{12}\cos k \\
     t_{12}\cos k & \sigma t_2\sin k
    \end{pmatrix}. \label{H_sigma}
\end{align}
The energy spectrum of the Hamiltonian (\ref{H_sigma}) is
\begin{equation}
    \epsilon_{k,\eta,\sigma} = \frac{t_2-t_1}{2}\sigma \sin k +\eta \sqrt{t_{12}^2\cos^2 k +\left(\frac{t_1+t_2}{2}\right)^2\sin^2k},
    \label{disp_general}
\end{equation}
where $\eta=\pm$ denotes the energy band. At $t_1=t_2=t_{12}=t$, the energy spectrum Eq.~\eqref{disp_general} becomes flat and spin independent, $\epsilon_{k,\eta,\sigma} = \eta t$, with the band gap being $2t$ (for definiteness, we assume $t>0$). The eigenstates of such a flat-band system are localized~\cite{Creutz1999}. We show the energy spectrum for flat and weakly dispersive bands in Fig.~\ref{fig:creutz_ladder-spectrum}.

Let us now introduce the interaction between fermions. We use the simplest Hubbard interaction. Since it is a local interaction, it does not couple electrons located on different ladder legs. Furthermore, it only couples opposite-spin electrons. The interacting Hamiltonian is given by
\begin{equation}
    \hat{H}_{\mathrm{int}} = \frac{U}{2}\sum_{j,\sigma} \left(a^\dagger_{j,\sigma}a_{j,\sigma}a^\dagger_{j,-\sigma}a_{j,-\sigma} +b^\dagger_{j,\sigma}b_{j,\sigma}b^\dagger_{j,-\sigma}b_{j,-\sigma}\right) \label{H_int_orig_local}
\end{equation}
with $U$ denoting the strength of the interaction. Substituting the local operators for the momentum operators, we get
\begin{align}
\hat{H}_{\mathrm{int}} = \frac{U}{2L} &\sum_{k, k', q,\sigma} \Big( a^\dagger_{k+q,\sigma}a^\dagger_{k'-q,-\sigma}a_{k',-\sigma} a_{k,\sigma} \nonumber\\
&+ b^\dagger_{k+q,\sigma}b^\dagger_{k'-q,-\sigma}b_{k',-\sigma} b_{k,\sigma} \Big). \label{H_int_orig}
\end{align}

As we show in Sec.~\ref{sec:quantum_metric}, the DQM and the GQM are formulated in terms of diagonal~\cite{Chen-vonGersdorff:2022} and original (nondiagonal)~\cite{Kashihara2023Mar} bases, respectively. The diagonal basis is defined as
\begin{equation}
    c_{k,\pm,\sigma} \equiv N_{k,\pm, \sigma} a_{k,\sigma} + M_{k,\pm, \sigma} b_{k,\sigma},
    \label{quasiparticle_def}
\end{equation}
where
\begin{subequations}
\label{N_k_and_M_k}
    \begin{align}
        N_{k,\pm, \sigma} &\equiv \sqrt{\frac{\epsilon_{k,\pm,\sigma} - \sigma t_2\sin k}{\epsilon_{k,\pm,\sigma} - \epsilon_{k,\mp,\sigma}}}, \label{N_k}\\
        M_{k,\pm, \sigma} &\equiv \frac{N_{k,\pm, \sigma} t_{12}\cos k}{\epsilon_{k,\pm,\sigma} -\sigma t_2 \sin k}. \label{M_k}
    \end{align}
\end{subequations}

Inverting Eq.~\eqref{quasiparticle_def} and substituting the result into Eq.\ \eqref{H_orig}, we obtain
\begin{equation}
    \hat{H} = \sum_{k,\eta,\sigma} \epsilon_{k,\eta,\sigma} c^\dagger_{k,\eta,\sigma} c_{k,\eta,\sigma}.
    \label{H_diag}
\end{equation}

To get the interaction in terms of the diagonal operators, we reexpress the operators $a_{k,\sigma}$ and $b_{k,\sigma}$ in terms of the diagonal operators $c_{k,\eta,\sigma}$. Substituting the result into Eq.~\eqref{H_int_orig}, we derive
\begin{align}
\label{H_int_diag}
&\hat{H}_{\mathrm{int}} =  \nonumber\\
&\frac{1}{L} \sum_{\substack{k, k', q,\sigma\\ \eta_1\dots\eta_4}} U_{k+q,k'-q,k',k}^{\eta_1 \eta_2 \eta_3 \eta_4,\sigma} c_{k+q, \eta_1, \sigma}^\dagger c_{k'-q, \eta_2, -\sigma}^\dagger c_{k', \eta_3, -\sigma} c_{k, \eta_4, \sigma},
\end{align}
where
%\begin{widetext}
\begin{align}
\label{H_int-U-def}
U_{k+q,k'-q,k',k}^{\eta_1 \eta_2 \eta_3 \eta_4,\sigma} &= \frac{U}{2} \Big(N_{k+q,\eta_1, \sigma}N_{k'-q,\eta_2, -\sigma}N_{k',\eta_3, -\sigma}N_{k,\eta_4, \sigma} \nonumber \\
&+M_{k+q,\eta_1, \sigma}M_{k'-q,\eta_2, -\sigma}M_{k',\eta_3, -\sigma}M_{k,\eta_4, \sigma} \Big)
\end{align}
%\end{widetext}
is the effective momentum- and spin-dependent interaction strength.

Each basis offers distinct advantages. While the diagonal basis offers a simple structure of the Hamiltonian $\hat{H}$ and, as we see below, the corresponding Green's functions, it leads to a more complicated interaction, see Eq.~\eqref{H_int-U-def}, coupling the two energy bands. Conversely, the interaction term is simple in the original basis. This basis, however, does not directly describe long-lived quasiparticles.

\subsection{Green's functions and self-energies}
\label{sec:creutz-G}

The Green's functions are crucial in defining both the DQM and the GQM. In what follows, we introduce the Green's functions in noninteracting and interacting systems. Interaction effects are taken into account via self-energy corrections. We treat interactions perturbatively and account only for the leading-order nontrivial terms in the interaction strength.

\subsubsection{Original basis}
\label{sec:creutz-G-orig}

In the original basis, the Green's function of a noninteracting system is block-diagonal in the spin space with each of the blocks defined as
\begin{equation}
\label{creutz-G-orig-def}
G^{\rm (orig)}_{0,\sigma}(k,i\omega_n) = \left[i\omega_n - H_{\sigma}(k)\right]^{-1},
\end{equation}
where $\omega_n=\pi T(2n+1)$ is the fermionic Matsubara frequency and $H_{\sigma}(k)$ is defined in Eq.~\eqref{H_sigma}. The Green's function can be conveniently represented as
\begin{equation}
\label{creutz-G-orig-1}
G^{\rm (orig)}_{0,\sigma}(k,i\omega_n) = \sum_{\eta=\pm} \frac{1}{i\omega_n-\epsilon_{k, \eta,\sigma}} C_{\eta, \sigma}(k),
\end{equation}
where
\begin{equation}
\label{creutz-G-orig-eta}
C_{\eta, \sigma}(k)
= \begin{pmatrix}
N_{k, \eta, \sigma}^2 & N_{k, \eta, \sigma} M_{k, \eta, \sigma}\\
N_{k, \eta, \sigma} M_{k, \eta, \sigma} & M_{k, \eta, \sigma}^2
\end{pmatrix}
\end{equation}
and the coefficients $N_{k, \eta, \sigma}$ and $M_{k, \eta, \sigma}$ are defined in Eq.~(\ref{N_k_and_M_k}).

The inverse interacting Green's function is
\begin{equation}
\label{creutz-G-orig-full}
\left[G^{\rm (orig)}(k,i\omega_n)\right]^{-1} = \left[G^{\rm (orig)}_{0}(k,i\omega_n)\right]^{-1} -\Sigma^{\rm (orig)}(k,i\omega_n),
\end{equation}
where $\Sigma^{\rm (orig)}(k,i\omega_n)$ is the self-energy matrix. We stress that the matrices in Eq.\ \eqref{creutz-G-orig-full} are matrices in both spin and band space. In the case of the Hubbard interaction, they are all block diagonal in spin, analogous to $H(k)$ in Eq.\ \eqref{H-k-spin}.

The leading nontrivial contribution to the self-energy is determined by the direct type of Feynman diagrams.
Other contributions either renormalize the Fermi energy or vanish, see, e.g., Ref.~\cite{abrikosov} for the corresponding diagrams. Such contributions are irrelevant for our analysis since we fix the renormalized Fermi energy. The details of these calculations are presented in Appendix ~\ref{app:self-energy}. Here, we state the final expression for the self-energy per spin projection,
\begin{widetext}
\begin{eqnarray}
\label{creutz-G-orig-Sigma}
\Sigma_{ab;\sigma}^{\rm (orig)}(k,i\omega_n)
&=& -U^2\sum_{\eta_1 =\pm} \sum_{\eta_2 =\pm} \sum_{\eta_3 =\pm} \int \frac{\mathrm{d} k'}{2\pi} \int \frac{\mathrm{d} k''}{2\pi} C_{\eta_1, ab, \sigma}(k+k') C_{\eta_2, ab, -\sigma}(k'') C_{\eta_3, ba, -\sigma}(k'+k'') \nonumber\\
&\times& F_{\eta_1,\eta_2,\eta_3}^{\rm (orig)}(k,k',k'')
=  \frac{U^2}{32} \begin{pmatrix}
4-\sigma \sin{k} & \cos{k} \\
\cos{k} & 4+\sigma \sin{k}
\end{pmatrix}_{ab} \sum_{\eta=\pm} \frac{1}{i\omega_n +3\eta t},
\end{eqnarray}
\end{widetext}
where $ab$ are ladder-leg indices and we defined
\begin{align}
\label{creutz-G-orig-F}
&F_{\eta_1,\eta_2,\eta_3}^{\rm (orig)}(k,k',k'')  \nonumber\\
&=\frac{\left[n_{\mathrm{B}}(\epsilon_1) +n_{\mathrm{F}}(\epsilon_{3}-\epsilon_2)\right] \left[n_{\mathrm{F}}(\epsilon_2) +n_{\mathrm{B}}(\epsilon_{3})\right]}{i\omega_{n} -\epsilon_1 -\epsilon_2 +\epsilon_3}\nonumber\\
&= -\sum_{\eta=\pm} \frac{\delta_{\eta_1,-\eta} \delta_{\eta_2,-\eta} \delta_{\eta_3,\eta}}{i\omega_{n} +3\eta t}
\end{align}
with $n_{\mathrm{F/B}}(\epsilon) = 1/\left(e^{\epsilon/T} \pm1\right)$ and used the shorthand notations $\epsilon_1 = \epsilon_{\eta_1, k+k', \sigma}$, $\epsilon_2 = \epsilon_{\eta_2, k'', \sigma}$, and $\epsilon_3 = \epsilon_{\eta_3, k'+k'', \sigma}$.
We transitioned from a summation over momentum to an integration, $(1/L)\sum_{k} \to \int \mathrm{d}k/(2\pi)$. In deriving the last expressions in Eqs.~\eqref{creutz-G-orig-Sigma} and \eqref{creutz-G-orig-F}, we assumed flat bands and that temperature is sufficiently low for the Fermi-Dirac and Bose-Einstein distributions to be accurately approximated by step functions.

The structure of the self-energy in the flat-band case, see Eq.~\eqref{creutz-G-orig-Sigma}, allows representing the interacting Green's function with the spin projection $\sigma$ \eqref{creutz-G-orig-full} in the following compact form:
\begin{align}
\label{creutz-G-orig-full-flat}
&\left[G_{\sigma}^{\rm (orig)}(k,i\omega_n)\right]^{-1} \nonumber\\
&= \Omega_U(i\omega_n) - t_U(i\omega_n) \begin{pmatrix}
-\sigma \sin{k} & \cos{k} \\
\cos{k} & \sigma \sin{k}
\end{pmatrix},
\end{align}
where
\begin{eqnarray}
\label{creutz-G-orig-OmegaU}
\Omega_U(i\omega_n)  &=& i\omega_n \left[1 -\frac{U^2}{4} \frac{1}{(i\omega_n)^2 -9t^2}\right],\\
\label{creutz-G-orig-tU}
t_U(i\omega_n)  &=& t\left[1 -\frac{3U^2}{16}\frac{1}{(i\omega_n)^2 -9t^2}\right].
\end{eqnarray}

Let us comment more on the physical meaning of the self-energy corrections in Eqs.~\eqref{creutz-G-orig-OmegaU} and \eqref{creutz-G-orig-tU}. Performing the analytical continuation to real frequencies as $i\omega_n \to \omega +i0^{+}$ and separating the real and imaginary parts of the self-energy via the Sokhotski–Plemelj formula, we find that the self-energy is peaked at $\omega=\pm 3t$ with its real and imaginary parts diverging as $1/(\omega \pm 3t)$ and $\df{\omega \pm 3t}$, respectively. This allows us to ignore the imaginary part for all $\omega$ not equal to $\pm 3t$. Then, the poles of the interacting Green's function follow from $\Omega_U^2(\omega)=t_U^2(\omega)$. This is a cubic equation in $\omega^2$ that has six roots $\pm \omega_{\alpha}$ with $\alpha=1,2,3$. In the leading order in $U/t$, the roots of $\Omega_U(\omega)=t_U(\omega)$ are
\begin{equation}
\label{creutz-G-omega-roots}
\omega_1 \approx 3t +\frac{3U^2}{64t}, \quad \omega_2 \approx t-\frac{U^2}{128t}, \quad \omega_3 \approx -3t -\frac{5U^2}{128t}.
\end{equation}
As we see, in addition to standard poles of the Green's function at $\omega =\pm t$, additional poles at $\omega = \pm 3 t$ appear. The latter originate exclusively from interaction-induced corrections, see the last terms in the square brackets in Eqs.~\eqref{creutz-G-orig-OmegaU} and \eqref{creutz-G-orig-tU}. These additional poles are similar to the replica bands studied in the context of superconductivity. There, the replica bands arise due to the frequency-dependence of the mass renormalization~\cite{Schrodi2021Replica} originating from the electron-phonon interaction, see also Refs.~\cite{Lee-Shen-InterfacialModeCoupling-2014, Rademaker-Johnston-EnhancedSuperconductivityDue-2016, Rebec-Shen-CoexistenceReplicaBands-2017, Aperis-Oppeneer-MultibandFullbandwidthAnisotropic-2018} for the discussions in the context of SrTiO$_3$ and FeSe, and Ref.~\cite{Chen2024} for the observation of replica bands in magic-angle twisted bilayer graphene.

In the original basis, the spectral function per spin projection is
\begin{align}
\label{creutz-G-orig-spectral}
&A_{\sigma}^{\rm (orig)}(k, \omega) \nonumber\\
&= -\frac{1}{2\pi i} \left[G_{\sigma}^{{\rm (orig)}}(k,\omega+i0^{+}) -G_{\sigma}^{{\rm (orig)}}(k,\omega -i0^{-}) \right] \nonumber \\
&= \mathrm{sgn}(\Omega_U(\omega))\df{\Omega^2_U(\omega)-t_U^2(\omega)} \nonumber \\
&\times\begin{pmatrix}
        \Omega_U(\omega) - \sigma t_U(\omega)\sin k & t_U(\omega) \cos k \\
        t_U(\omega) \cos k & \Omega_U(\omega) + \sigma t_U(\omega)\sin k
    \end{pmatrix},
\end{align}
where the analytical continuation to real frequencies was performed in the standard way, $i\omega_n \to \omega \pm i0^{\pm}$ with the sign $+$ ($-$) corresponding to the retarded (advanced) Green's function. The last equality applies to the flat-band case at low temperature. Since the imaginary part of the self-energy is peaked at $\omega=\pm 3t$, it does not affect the spectral function determined by the poles of the interacting Green's function.~\footnote{As follows from Eq.~\eqref{creutz-G-omega-roots}, the poles of the interacting Green's function are always away from $\pm 3t$. Therefore, taking into account that $\IM{\Sigma^{\rm (orig)}} \propto \df{\omega \pm 3t}$, the spectral function vanishes at $\omega=\pm3t$.} As we will show in Sec.~\ref{sec:disp}, this is generally not the case for dispersive bands where both real and imaginary parts of the self-energy should be taken into account.

\subsubsection{Diagonal basis}
\label{sec:creutz-G-diag}

The Green's function for the band $\eta$ and the spin projection $\sigma$ in the diagonal basis is
\begin{equation}
\label{creutz-G-diag-eta}
G^{\rm (diag)}_{0,\eta,\sigma}(k,i\omega_n) =  \frac{1}{i\omega_n-\epsilon_{k,\eta,\sigma}}.
\end{equation}

The interacting Green's function is defined similarly to that in the original basis, see Eq.~\eqref{creutz-G-orig-full}. Referring to Appendix \ref{app:self-energy} for details, the final expression for the self-energy reads
\begin{eqnarray}
\label{creutz-G-diag-Sigma-def}
\Sigma_{\eta\eta'',\sigma}^{\rm (diag)}(k,i\omega_n) &=& -4 \sum_{\eta_1 \eta_2 \eta'} \int \frac{\mathrm{d}k'}{2\pi} \int \frac{\mathrm{d}q}{2\pi} U_{k+q, k'-q, k', k}^{\eta_1 \eta_2 \eta' \eta'',\sigma} \nonumber\\
&\times& U_{k', k, k+q, k'-q}^{\eta' \eta \eta_1 \eta_2,-\sigma}
F^{\rm (diag)}_{\eta',\eta_1,\eta_2}(k,k',q) \nonumber\\
&=& \frac{U^2}{16} \delta_{\eta\eta''} \frac{4i\omega_n -3\eta t}{(i\omega_n)^2 -9t^2},
\end{eqnarray}
where
\begin{align}
\label{creutz-G-diag-F}
&F^{\rm (diag)}_{\eta',\eta_1,\eta_2}(k,k',q) = \nonumber\\
&-\frac{[n_F(\epsilon_1)+n_B(\epsilon_3)][n_F(\epsilon_1-\epsilon_3)+n_B(\epsilon_2)]}{i\omega_n +\epsilon_1-\epsilon_2-\epsilon_3} \nonumber\\
&=- \sum_{\eta=\pm} \frac{\delta_{\eta',\eta} \delta_{\eta_1,-\eta} \delta_{\eta_2,-\eta}}{i\omega_n +3\eta t}
\end{align}
with $\epsilon_1 = \epsilon_{k'\eta',-\sigma}$, $\epsilon_2 = \epsilon_{k+q,\eta_1,\sigma}$, and $\epsilon_3 = \epsilon_{k'-q,\eta_2,-\sigma}$. In the last expressions in Eqs.~\eqref{creutz-G-diag-Sigma-def} and \eqref{creutz-G-diag-F} we assumed the flat-band and low-temperature limits.

Similar to the original basis, the structure of the self-energy \eqref{creutz-G-diag-Sigma-def} makes it possible to present the interacting Green's function in a compact form for the flat-band case
\begin{equation}
\label{creutz-G-diag-full-flat}
G_{\eta, \sigma}^{\rm (diag)}(k,i\omega_n)= \frac{1}{\Omega_U(i\omega_n) - \eta t_U(i\omega_n)}.
\end{equation}
The full Green's function matrix $G^{\rm (diag)}(k,i\omega_n)$ is diagonal in band space and spin-independent. The spectral function in the diagonal basis inherits this form
\begin{align}
    A_{\eta, \sigma}^{\rm (diag)}(k, \omega) = \df{\Omega_U(\omega) - \eta t_U(\omega)}.
    \label{spectral_diagonal}
\end{align}
Note that the poles in the interacting Green's functions and, as a result, the peaks in the spectral function are the same in both bases, as they should be.

We close this section with a more detailed discussion of replica bands and the imaginary part of the self-energy. Replica bands arise due to oscillations of the real part of the self-energy in some frequency range, giving more peaks in the spectral function. From causality \cite{Toll1956Causality}, the real part of the self-energy is related to the imaginary part via the Kramers-Kronig relations revealing valuable physical insights just by studying the imaginary part alone. The sharply peaked density of states associated with flat bands results in sharp peaks in the imaginary part of the self-energy. These sharp peaks give large and rapid oscillations of the real part in the same frequency range. In this case, the replica bands appear close to $\omega = \pm 3t$ because the imaginary part of the self-energy has a $\delta$-function peak there. The appearance of the latter follows from the conservation of energy and momentum as we discuss below.

Originally, we have doubly degenerate bands at $\pm t$. At low enough temperature, we can think of the lower bands as completely filled, and the upper bands as completely empty. Our interaction can alter band indices or leave them unchanged, while the spin indices are always conserved, see Eq.~\eqref{H_int_diag}. Hence, interactions can change electron energy by $0$ or $2t$. When $\omega>0$, a virtual electron can be created with energy $\omega$. Then, $|\IM\Sigma(\omega)|$ is the rate with which this electron relaxes to an available on-shell electron state due to an interaction. By considering energy conservation in the self-energy diagram in Fig.~\ref{fig:self-energy-diag} and forcing the internal lines to have energies $\pm t$, we find that scattering is possible if the incoming electron has energy $3t$. Then a virtual electron with energy $3t$ interacts with a filled electron state in $-t$ to generate two electrons with energy $t$. Hence, scattering can occur and $\IM\Sigma(\omega = 3t) \neq 0$. On the other hand, $\IM\Sigma(\omega = t) = 0$ because a virtual electron would have to interact with an electron from the filled band, and there is no intraband scattering in a filled band. Analogous discussions apply for $\omega < 0$.

In the case of dispersive bands, the dispersion opens up a wider range of possible energy transfers, and the $\delta$-function contributions to $\IM\Sigma(\omega)$ at $\omega = \pm 3t$ become extended peaks in a range around $\omega = \pm 3t$. In that sense, even weakly dispersive bands give fundamentally different effects compared with exactly flat bands.

\section{Quantum metric}
\label{sec:quantum_metric}

In this section, we introduce the generalized and dressed quantum metric as well as describe a numerical approach based on the exact diagonalization. The exact diagonalization provides an important benchmark for the other two approaches to the quantum metric, although it is limited to relatively small systems and momentum-averaged quantum metric. While the main focus of this section is on flat-band systems, we also give general expressions for systems with dispersive bands.

\subsection{Generalized quantum metric}
\label{sec:quantum_metric-GQM}

The GQM~\cite{Kashihara2023Mar} is based on the generalization of the noninteracting expressions of Ref.~\cite{Souza-Martin:2000} where the quantum metric $g(k)$ satisfies the following relation:
\begin{equation}
g = \int \frac{\mathrm{d}k}{2\pi} g(k) = \int_0^\infty\mathrm{d}\Omega \frac{\RE{\sigma(\Omega)}}{\pi\Omega}.
    \label{metric_cond_relation}
\end{equation}
Here $\sigma(\Omega)$ is the optical conductivity. By using the Kubo formula, the latter can be represented as
\begin{equation}
    \RE{\sigma(\Omega)} = \frac{1}{\Omega} \IM{\Pi(\Omega+i0^{+})}
    \label{sigma_polar_relation}
\end{equation}
with the current-current correlation function
$\Pi(\Omega+i0^{+})$ being
\begin{align}
&\Pi(\Omega+i0^{+}) = -\int_{-\infty}^{\infty}\int_{-\infty}^{\infty} \mathrm{d}\omega\mathrm{d}\omega' \nonumber \frac{n_{\mathrm{F}}(\omega)-n_{\mathrm{F}}(\omega')}{\omega-\omega'-\Omega-i0^{+}} \\
&\times\int\frac{\mathrm{d}k}{2\pi} \mathrm{tr}\big[J(k) A^{\rm (orig)}(k, \omega) J(k) A^{\rm (orig)}(k,\omega') \big],
    \label{Pi_def}
\end{align}
see, e.g., Ref.~\cite{Mahan:book-2013}. Here, the trace runs over the band and spin spaces, $J(k)$ is the current operator $J(k)=\partial H(k)/\partial k$, and $A^{\rm (orig)}(k, \omega)$ is the interacting spectral function in the original basis defined in Eq.~\eqref{creutz-G-orig-spectral}. Note that the expression in Eq.~\eqref{Pi_def} can be rewritten in the diagonal basis, albeit it becomes less transparent in the interacting case, see Ref.~\cite{Kashihara2023Mar}.

Using Eqs.~\eqref{metric_cond_relation}, \eqref{sigma_polar_relation}, and \eqref{Pi_def}, the momentum-dependent quantum metric reads
\begin{align}
g^{\rm GQM}(k) &= -\int_0^\infty \frac{\mathrm{d}\Omega}{\Omega^2} \int_{-\infty}^{\infty}\mathrm{d}\omega \nonumber \left[n_{\mathrm{F}}(\omega)-n_{\mathrm{F}}(\omega-\Omega)\right] \nonumber\\
&\times
\mathrm{tr}\big[J(k) A^{\rm (orig)}(k, \omega) J(k) A^{\rm (orig)}(k,\omega-\Omega) \big].
    \label{metric_cond_relation-gk}
\end{align}
This expression is suitable for both analytical and numerical calculations.

In the flat-band limit, the quantum metric can be calculated analytically, while we resort to numerical methods in the case of dispersive bands in Sec.~\ref{sec:disp}. We leave the details of the calculations to Appendix~\ref{app:g_steps} and present the final result for the quantum metric in the flat-band and zero-temperature limits
\begin{eqnarray}
\label{GQM-g-fin}
g^{\rm GQM} &=&  \sum_{\alpha,\beta} \frac{2t^2}{(\omega_{\alpha}+\omega_{\beta})^2} \nonumber\\
&\times& \frac{\sign{\omega_{\alpha} +\omega_{\beta}} \sign{\omega_{\alpha}}}{ \left|\Omega_{U}'(\omega_{\alpha}) - t_{U}'(\omega_{\alpha})\right| \left|\Omega_{U}'(\omega_{\beta}) - t_{U}'(\omega_{\beta})\right|} \nonumber\\
&\approx& \frac{1}{2} -\frac{5U^2}{256t^2}.
\end{eqnarray}
Here, we used Eq.~\eqref{creutz-G-omega-roots} for the roots $\omega_{\alpha}$ with $\alpha=1,2,3$ and expanded in $U/t\ll1$ in the last expression. Interaction effects thus suppress the quantum metric compared with its noninteracting value $1/2$. In addition, the quantum metric does not depend on momentum, $g^{\rm GQM}(k)=g^{\rm GQM}$. Before discussing Eq.~\eqref{GQM-g-fin} further and comparing it with the exact diagonalization results, we calculate the quantum metric via a different approach dubbed the DQM and advocated in Ref.~\cite{Chen-vonGersdorff:2022}.

\subsection{Dressed quantum metric}
\label{sec:quantum_metric-DQM}

The DQM~\cite{Chen-vonGersdorff:2022} originates from the generalization to the interacting case of the fidelity susceptibility, see, e.g., Refs.~\cite{You-Gu-FidelityDynamicStructure-2007, Zanardi-Cozzini-InformationTheoreticDifferentialGeometry-2007, Gu-FidelityApproachQuantum-2010} and reads as
\begin{align}
    g^{\rm DQM}(k) &= \sum_{\eta_1,\eta_2} \int_{0}^\infty \mathrm{d}\Omega \int_{-\infty}^\infty \mathrm{d}\omega \left[n_{\mathrm{F}}(\omega) - n_{\mathrm{F}}(\omega + \Omega)\right]\nonumber \\
    &\times |\mathcal{A}_{\eta_1\eta_2}(k)|^2 A_{\eta_1}^{\rm (diag)}(k,\omega) A_{\eta_2}^{\rm (diag)}(k,\omega+\Omega) .
    \label{g_def_CvG}
\end{align}
Here, the Berry connection $\mathcal{A}_{\eta_1\eta_2}(k)$ is
\begin{equation}
    \mathcal{A}_{\eta_1\eta_2}(k) = \langle k,\eta_1|i\partial_k|k,\eta_2\rangle
    \label{berry_curv_def}
\end{equation}
and the spectral function $A_{\eta}^{\rm (diag)}(k,\omega)$ in the diagonal basis is given in Eq.~\eqref{spectral_diagonal}. Above, $|k,\eta\rangle$ is the single-particle eigenstate of band $\eta$.
For the model at hand, only the off-diagonal components of the Berry connection are nonzero which in the flat-band limit read as
\begin{equation}
    \mathcal{A}_{+-}(k) = -\mathcal{A}_{-+}(k) = \frac{i}{2}\mathrm{sgn}(\cos(k)).
    \label{berry_curv_solved}
\end{equation}
Thus, for all $k$, we have $|\mathcal{A}_{+-}(k)|^2= 1/4$.

As in the case of the GQM discussed in Sec.~\ref{sec:quantum_metric-GQM}, integrations in Eq.~\eqref{g_def_CvG} are performed analytically in the flat-band regime. Referring to Appendix~\ref{app:g_steps} for details, one obtains the following compact expression in the zero-temperature limit:
\begin{eqnarray}
g^{\rm DQM} &=& \frac{1}{2}\sum_{\alpha,\beta} \frac{\mathrm{sgn}(\omega_{\alpha} + \omega_{\beta}) \mathrm{sgn}(\omega_{\alpha}) }{|\Omega_U'(\omega_{\alpha}) - t_U'(\omega_{\alpha})||\Omega_U'(\omega_{\beta}) - t_U'(\omega_{\beta})|} \nonumber\\
&\approx& \frac{1}{2} -\frac{5U^2}{512t^2},
\label{g_k_cvg_solved}
\end{eqnarray}
where we used Eq.~\eqref{creutz-G-omega-roots} and expanded in $U/t\ll1$ in the last expression. Like the GQM, the DQM is momentum independent in the flat-band case, $g^{\rm DQM}(k) =g^{\rm DQM}$.

The two approaches to calculating the quantum metric yield similar, but distinct, analytical expressions, cf. Eqs.\ \eqref{GQM-g-fin} and \eqref{g_k_cvg_solved}. They coincide in the noninteracting limit, as both reduce to the conventional expression for $g(k)$ \cite{Chen-vonGersdorff:2022, Kashihara2023Mar}, but differ in the presence of interactions. By denoting the contributions to $g$ from the poles at $\pm \omega_{\alpha}$ and $\pm \omega_{\beta}$ as $g_{\alpha \beta}$ for the two methods, we can quantify the difference between them as
\begin{equation}
    \frac{g_{\alpha \beta}^{\mathrm{DQM}}}{g_{\alpha \beta}^{\mathrm{GQM}}} = \frac{(\omega_{\alpha} +\omega_{\beta})^2}{4t^2}.
    \label{ratio_disc}
\end{equation}
At $U\to0$, there is only one pole $\omega_\alpha=t$, such that $\omega_{\alpha}+\omega_{\beta} = 2t$ leading to $g_{\alpha \beta}^{\mathrm{DQM}}=g_{\alpha \beta}^{\mathrm{GQM}}$. The difference between the roots $\omega_{\alpha}$ in the noninteracting and interacting system results in the difference between the DQM and the GQM at $U\neq0$. Intuitively, a stronger dependence on the interaction strength in the GQM may be linked to its less direct definition which involves additional $\Omega$-dependent factors affected by interactions. Since interaction effects reduce the quantum metric, their overestimation in the GQM results in a stronger dependence on $U/t$. We view the factor of 2 difference between the prefactor of the $U^2/t^2$ term in Eqs.~\eqref{GQM-g-fin} and \eqref{g_k_cvg_solved} to be model specific, while the fact that they are different is more general.

\subsection{Many-body quantum metric}
\label{sec:quantum_metric-MBQM}

Given the differences between the GQM and the DQM, it is worth assessing which is more accurate. We use exact diagonalization to obtain the many-body ground state for the interacting Creutz ladder. This allows us to directly compare both methods against exact results for small system sizes.

The many-body quantum metric (MBQM) in a periodic 1D system can be defined in terms of the flux $\Phi$ threading the system~\cite{Provost1980a, Souza-Martin:2000, Salerno-Torma:2023}. Introducing, $\phi=\Phi/L$, we have
\begin{equation}
    g^{\mathrm{MBQM}}(\phi) = L \langle\partial_\phi\Psi(\phi)|\partial_\phi\Psi(\phi)\rangle - L | \langle\partial_\phi \Psi(\phi)| \Psi(\phi)\rangle|^2,
    \label{g_MB_def}
\end{equation}
where $|\Psi(\phi)\rangle$ is the many-body ground state and, to ease the comparison with the DQM and the GQM, we multiplied the standard expression for the MBQM by $L$. %For small $\phi$, the
The derivative of the ground state in the zero-flux limit $\phi \to 0$ is defined as
\begin{equation}
    |\partial_\phi \Psi(0)\rangle = \lim_{\phi\to 0}\frac{|\Psi(\phi)\rangle-|\Psi(0)\rangle}{\phi}.
\end{equation}
By using this relation in Eq.~\eqref{g_MB_def}, we obtain
\begin{equation}
    g^{\mathrm{MBQM}} \equiv g^{\mathrm{MBQM}}(0) = L\lim_{\phi\to0} \frac{1 - |\langle \Psi(\phi)|\Psi(0)\rangle|^2}{\phi^2}.
    \label{g_mb_calc}
\end{equation}

In the noninteracting regime, the quantum metric in momentum space, $g(k)$, is directly related to the MBQM given in Eq.\ \eqref{g_MB_def} and evaluated for the ground state in the zero-flux limit $\phi \to 0$ as~\cite{Souza-Martin:2000, Ozawa2019Nov}
\begin{equation}
    g^{\mathrm{MBQM}} = \frac{1}{L}\sum_k g(k).
    \label{many_body_single_particle_metric_relation}
\end{equation}
While $g^{\mathrm{MBQM}}$ can only be calculated for small systems, it is exact, also in the presence of interactions. Since Eq.~\eqref{many_body_single_particle_metric_relation} holds for noninteracting systems~\footnote{In noninteracting cases, the momentum sum over the quantum metric has also been connected to the spread of Wannier functions, see, e.g., Refs.~\cite{Marzari-Vanderbilt-MaximallyLocalizedGeneralized-1997, Marzari2012WannierRev, deSousa2023Chen1, Cardenas-Castillo2024Chen2}.} and $g(k)$ is calculated perturbatively, we do not expect Eq.\ \eqref{many_body_single_particle_metric_relation} to hold exactly when interactions are included. Nevertheless, $g^{\mathrm{MBQM}}$ serves as a useful benchmark for our analytical results, allowing us to compare the GQM and the DQM from the previous sections with the MBQM.

To calculate the MBQM given in Eq.\ \eqref{g_mb_calc}, we use the same method as in Ref.\ \cite{Salerno-Torma:2023}, which considered the bosonic Creutz ladder. We use $\hat{H}$ and $\hat{H}_{\mathrm{int}}$ as the single-particle and interacting Hamiltonians, respectively. We fix the system to be at half-filling, meaning that there are $2L$ electrons: $L$ with spin up and $L$ with spin down. This ensures correspondence with our analytical calculations where the Fermi energy is placed inside the band gap, such that the system is always at half filling.~\footnote{Note that exact diagonalization uses a canonical ensemble where particle number is fixed. The Green's function approach uses a grand canonical ensemble. To make a valid comparison, we fixed the renormalized Fermi energy to ensure half-filling also in the Green's function approach.}
To account for the flux threading the ring, all forward (backward) hoppings acquire a phase $\mathrm{e}^{i\phi}$ ($\mathrm{e}^{-i\phi}$), such that traversing the entire ring acquires a global phase $\mathrm{e}^{\pm i\phi}$. With the Hamiltonian in place, we use the \texttt{QuSpin} exact diagonalization package \cite{Weinberg2017, Weinberg2019} to obtain the many-body ground state. In this way, we calculate $|\Psi(\phi)\rangle$ for $\phi=0$ and for a sufficiently small value of $\phi$. Finally, we use these two states in Eq.~\eqref{g_mb_calc} to compute the left-hand side of Eq.~\eqref{many_body_single_particle_metric_relation}, allowing comparison with the analytical expressions in Eqs.~\eqref{GQM-g-fin} and \eqref{g_k_cvg_solved}. When comparing, we always use results from our largest system, namely, $L=6$.

\section{Results and discussion}
\label{sec:results}

\subsection{Flat bands}

\begin{figure}[ht]
    \centering
    \includegraphics[width=\columnwidth]{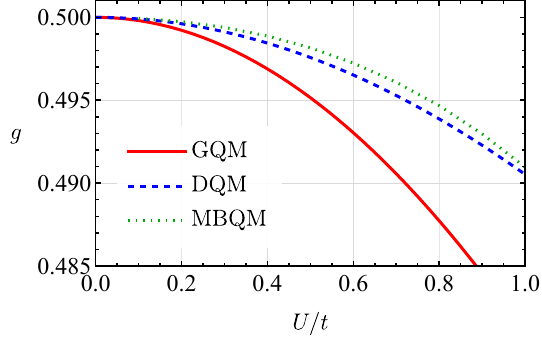}
    \includegraphics[width=\columnwidth]{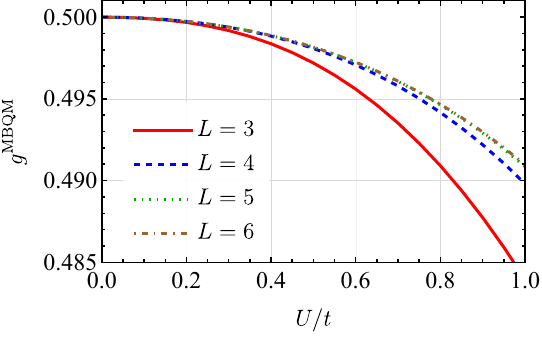}
    \caption{
    (Top panel) Comparison of quantum metric in three different approaches. The solid red line corresponds to the GQM, the dashed blue line denotes the DQM, and the dotted green line corresponds to the MBQM. (Bottom panel) The many-body quantum metric $g^{\mathrm{MBQM}}$ calculated using exact diagonalization in an interacting fermionic Creutz ladder. It is plotted as a function of the strength of the Hubbard interaction $U$ for different values of $L$. In both panels, we assume the flat-band limit $t_1=t_2=t_{12}=t$.
    }
    \label{fig:g_comp_all_three_methods}
\end{figure}

For the case of flat bands, we compare the results for the GQM and the DQM with the MBQM in the top panel of Fig.~\ref{fig:g_comp_all_three_methods}. As one can see, the DQM shows a good agreement with the MBQM, while the dependence of the GQM on the interaction strength is too strong. This comparison establishes the DQM as a more precise way to evaluate the quantum metric in the interacting Creutz ladder.

The GQM and the DQM are formulated for infinite systems, while the MBQM can only be calculated for a finite one, so let us briefly consider how $g^{\mathrm{MBQM}}$ converges as a function of the ladder length $L$. In the bottom panel of Fig.\ \ref{fig:g_comp_all_three_methods}, we plot $g^{\mathrm{MBQM}}$ as a function of $U$ for different choices of $L$, allowing us to assess the impact of finite-size effects. Since we use a perturbative approach to the GQM and the DQM, we consider the interval of $U=[0,1]t$; the case with $U>t$ is considered in the exact diagonalization approach in Sec.~\ref{sec:strong}. In the noninteracting limit, $g^{\mathrm{MBQM}} \rightarrow 1/2$ regardless of the length of the lattice, in agreement with the analytical results. As the interaction increases, the choice of $L$ has a stronger effect on $g^{\mathrm{MBQM}}$, especially at low values of $L$, see the bottom panel in Fig.~\ref{fig:g_comp_all_three_methods}. However, $g^{\mathrm{MBQM}}$ still converges with increasing $L$, as the difference between $L=5$ and $L=6$ is minimal. This indicates that $g^{\mathrm{MBQM}}$ for $L = 6$ offers a reliable approximation of bulk behavior in the flat-band case, allowing for a meaningful comparison of the MBQM with the GQM and the DQM. The similarity between small and large systems can be attributed to the localized nature of the eigenstates in the noninteracting case, making the system less sensitive to the addition of more unit cells.

\subsection{Weakly dispersive systems}
\label{sec:disp}

So far, we demonstrated that the DQM provides a better fit than the GQM to the MBQM obtained via exact diagonalization in the flat-band limit. We now establish that this result is not merely an artifact of exactly flat bands. We use the model in Eq.~\eqref{H_sigma}, but relax the flat-band assumption by setting $t_{12}\neq t_1, t_2$. The bands are assumed to be weakly dispersive with a bandwidth much smaller than the band gap. While allowing us to simplify the computations, such bands are still fundamentally different from the flat bands.

The momentum-dependent quantum metric is defined in Eqs.~\eqref{metric_cond_relation-gk} and \eqref{g_def_CvG} for the GQM and the DQM, respectively. The integrals over frequencies are calculated numerically where we also introduced broadening in the spectral functions as $\omega \to \omega +i\delta$. We use the self-energies calculated via the first expressions in Eqs.~\eqref{creutz-G-orig-Sigma} and \eqref{creutz-G-diag-Sigma-def}; we obtain the imaginary part first and use the Kramers-Kronig relation to derive the real part.

\begin{figure}[t]
    \centering
    \includegraphics[width=0.95\columnwidth]{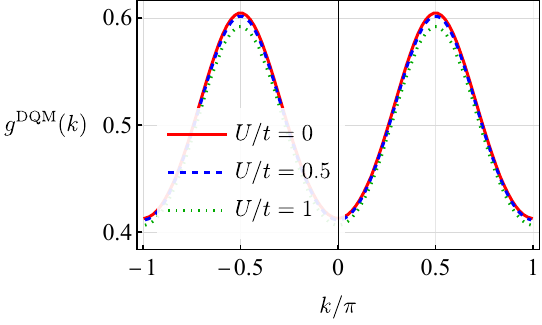}
    \caption{Dependence of the DQM on momentum for a few values of interaction strength $U/t$. We set $t_1=t_2=t$ and $t_{12}=1.1\,t$.}
    \label{fig:disp-g-k}
\end{figure}

We show the DQM in Fig.~\ref{fig:disp-g-k} for a few values of $U/t$. As one can see, the quantum metric has a strong dependence on $k$ but shows only minute variations with the interaction strength $U/t$.

\begin{figure}[t]
    \centering
    \includegraphics[width=\columnwidth]{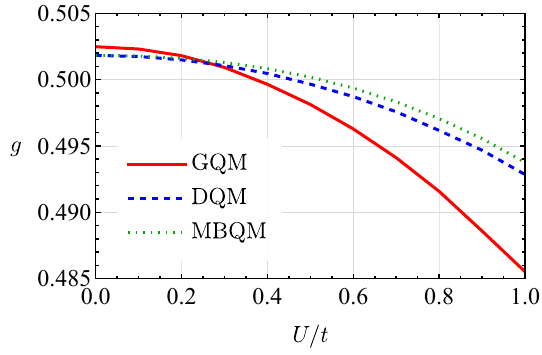}
    \includegraphics[width=\columnwidth]{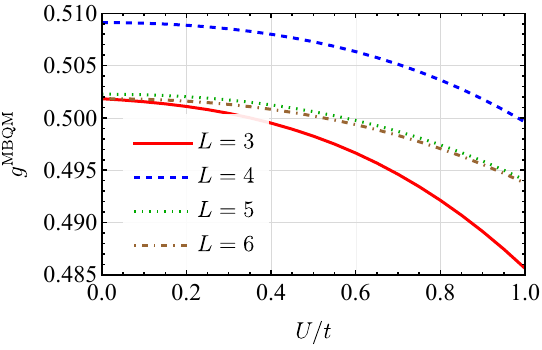}
    \caption{Same as in Fig.~\ref{fig:g_comp_all_three_methods} but for weakly dispersive bands with $t_1=t_2=t$ and $t_{12}=1.1\,t$. In calculating the integrals over frequencies, we replaced $\omega \to \omega +i\delta$ with $\delta=10^{-3}\,t$ in the spectral functions. The discrepancy between the GQM and the DQM at $U/t\to0$ arises due to more demanding and less precise numerical calculations necessary for dispersive bands.
    }
    \label{fig:disp-g}
\end{figure}

To compare the momentum-dependent generalized and dressed quantum metrics with the MBQM, these expressions are integrated over momentum. We show the quantum metrics for weakly dispersive bands in the top panel of Fig.~\ref{fig:disp-g}. As in the case of flat bands, the DQM shows great agreement with the MBQM, especially at small values of $U/t$. The minute discrepancy between the DQM and the GQM at $U/t\to0$ is an artifact of more demanding numerical integrations over frequencies necessary for dispersive bands. Thus, deviations from the flat-band case, i.e., a different dispersion relation, do not affect the agreement between the DQM and the MBQM supporting the generality of our statement that the DQM provides a better fit to the MBQM compared with the GQM.

A good agreement between the DQM and the MBQM may have its roots in a direct relation between the fidelity susceptibility and the DQM. This makes the DQM a more straightforward definition of the quantum metric in interacting systems. Indeed, the fidelity susceptibility is directly linked to the overlap of the wave functions and is well-defined even for interacting systems~\cite{You-Gu-FidelityDynamicStructure-2007, Zanardi-Cozzini-InformationTheoreticDifferentialGeometry-2007, Gu-FidelityApproachQuantum-2010}. To our knowledge, a similar connection between the optical conductivity, used to define the GQM, and the overlap of wavefunctions does not exist in the interacting case.

In experimental settings, interactions are rarely strictly zero, meaning that the measured quantum metric inherently contains interaction effects. Our findings that the DQM fits well with the results from exact diagonalization indicate that the measurement methods suggested in Ref.~\cite{Chen-vonGersdorff:2022} provide a viable addition to the methods used in Refs.~\cite{Tan-Yu-ExperimentalMeasurementQuantum-2019, Yu-Cai-ExperimentalMeasurementQuantum-2020, Gianfrate-Malpuech-MeasurementQuantumGeometric-2020, Kang-Comin-MeasurementsQuantumGeometric-2024}. Moreover, with the DQM, one obtains the quantum metric as a function of $k$, which is experimentally accessible \cite{Kang-Comin-MeasurementsQuantumGeometric-2024}. The MBQM, on the other hand, does not allow probing the quantum metric for any specific $k$, only the sum over $k$, see Eq.\ \eqref{many_body_single_particle_metric_relation}.

Note that the MBQM for the dispersive bands is more sensitive to the size of the system, see the bottom panel of Fig.~\ref{fig:disp-g}. So, while the results for $L=5$ and $L=6$ show good convergence, for dispersive systems it is possible to attribute some of the discrepancies between the quantum metrics to finite-size effects in the MBQM.

\subsection{Strong-coupling regime}
\label{sec:strong}

\begin{figure}
    \centering
    \includegraphics[width=\linewidth]{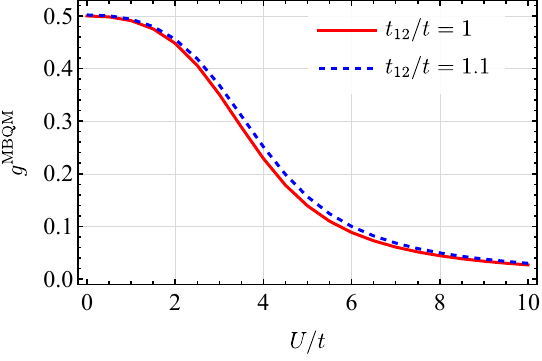}
    \caption{The many-body quantum metric $g^{\mathrm{MBQM}}$ calculated for the interacting fermionic Creutz ladder for $L=6$ using exact diagonalization, as a function of the Hubbard interaction $U$.
    We fix $t_1=t_2=t$ and set $t_{12}=t$ and $t_{12}=1.1\,t$ in the flat-band and dispersive cases, respectively.
    }
    \label{fig:strong_coupling}
\end{figure}

For completeness, we also include exact diagonalization results for the strong-coupling regime. First, in Fig.\ \ref{fig:strong_coupling}, we plot $g^{\mathrm{MBQM}}$ for $L=6$ for the same systems as in Figs.\ \ref{fig:g_comp_all_three_methods} and \ref{fig:disp-g}. We observe that in both flat and dispersive cases, $g^{\mathrm{MBQM}}$ continues to decay monotonically with $U/t$. This decay saturates; $g^{\mathrm{MBQM}}$ remains positive for large values of $U$, as any metric should be.

The saturation of the quantum metric with the interaction strength has a transparent physical meaning. Indeed, as was established for noninteracting systems in Ref.~\cite{Souza-Martin:2000}, the quantum metric is directly related to the localization length $\xi$, i.e., $\xi^2\propto g$. Thus, the states of an interacting system become increasingly more localized as $U/t\to\infty$.

\subsection{Spectral properties}
\label{sec:spectral}

We close this section by considering the spectral properties of the interacting fermionic Creutz ladder. Spectral probes such as the spectral function and the optical conductivity provide a direct way to investigate the band structure and interaction effects. As we explicitly showed for the flat-band system, interactions shift the poles of the Green's function and allow for additional replica bands, see, e.g., Eq.~\eqref{creutz-G-omega-roots}. Therefore, the optical conductivity and the spectral function provide another glimpse into the interaction effects.

For flat-band systems, the spectral function in the original and diagonal bases is given in Eqs.~\eqref{creutz-G-orig-spectral} and \eqref{spectral_diagonal}, respectively. We focus on the trace of the spectral function, which is the same for both bases and reads
\begin{eqnarray}
\label{spectral-A}
\mbox{tr}\,A(k, \omega) &=& 4\left|\Omega_U(\omega)\right| \df{\Omega^2_U(\omega)-t_U^2(\omega)} \nonumber\\
&=& 2\sum_{\alpha=1,2,3} \frac{\df{\omega -\omega_{\alpha}}}{\left|\Omega_{U}'(\omega_{\alpha}) - t_{U}'(\omega_{\alpha})\right|}.
\end{eqnarray}
According to Eq.~\eqref{creutz-G-omega-roots}, the flat-band spectral function is peaked at $\omega \approx t$ and $\omega \approx 3t$ with the last peak being due to replica bands. We note that for flat bands, while the interactions shift the bands, the width of the peaks in Eq.~\eqref{spectral-A} remains unaffected. Thus, regardless of their strength, the interaction effects do not lead to any broadening of the flat bands.

\begin{figure}
    \centering
    \includegraphics[width=\columnwidth]{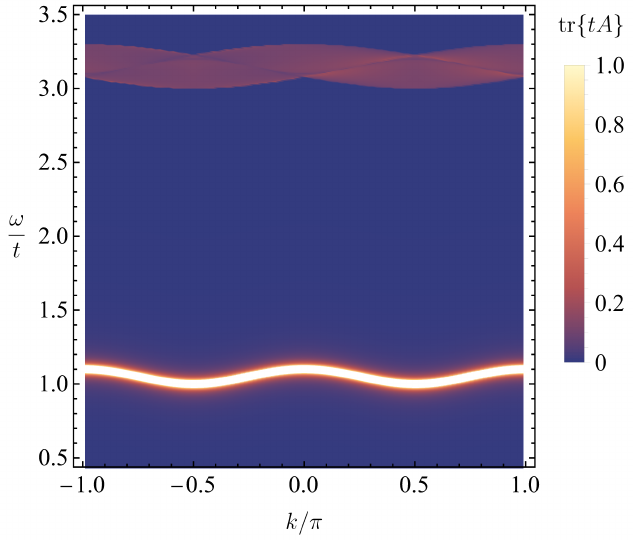}
    \caption{The trace of the spectral function \eqref{spectral-A} at $U=0.75\,t$. We fix $t_1=t_2=t$ and $t_{12}=1.1\,t$. To visualize the original band at $\omega \sim t$, we included the artificial broadening $\omega \to \omega +i\delta$ with $\delta=10^{-3}\,t$.}
    \label{fig:spectral-A}
\end{figure}

Away from the flat-band limit, one should use the spectral function defined in the first line in Eq.~\eqref{creutz-G-orig-spectral}. In Fig.~\ref{fig:spectral-A}, we show the spectral function Eq. \eqref{spectral-A} for the weakly dispersive bands, where one can observe the broadening of the replica bands introduced by interactions. We deliberately choose a large value of $U/t$ to emphasize the replica band appearing at $\omega\sim 3t$. Note also that the replica band has a more complicated structure with a nonuniform intensity compared with the original band. While the intensity of the original band is determined by the artificial broadening parameter that we include as $\omega \to \omega +i\delta$, the intensity of the replica band is strongly affected by the imaginary part of the self-energy.

Being a measurable quantity, the optical conductivity $\sigma(\Omega)$ provides another way to directly probe interaction effects. Using Eq.~\eqref{creutz-G-omega-roots} in Eqs.~\eqref{sigma_polar_relation} and \eqref{Pi_def} and expanding up to the second order in $U/t$, see Eq.~\eqref{g_steps-GQM-sigma} for the explicit expression for the optical conductivity, we obtain in the flat-band limit
\begin{eqnarray}
\label{analytics-sigma-expl}
\RE{\sigma(\Omega)} &\approx& \frac{2\pi t^2}{\Omega} \Bigg[\left(1-\frac{17}{256} \frac{U^2}{t^2}\right) \df{2t -\frac{U^2}{64t^2} -\Omega} \nonumber\\
&+&\frac{3}{64} \frac{U^2}{t^2} \df{4t-\Omega}
\Bigg].
\end{eqnarray}
First, there are transitions involving only the original bands at $\pm t$, see the first term in the square brackets in Eq.~\eqref{analytics-sigma-expl}, and both original and interaction-induced replica bands at $\pm 3t$, see the second term in the square brackets in Eq.~\eqref{analytics-sigma-expl}. Transitions between only replica bands are of the fourth order in $U/t$ and, hence, are beyond the range of applicability of our theory. Second, the peak of the optical conductivity due to transitions between the original bands is shifted due to interaction effects. Thus, the interaction effects are manifested in both the renormalization of the features present in a noninteracting system and the appearance of the new peaks due to the replica bands.

\section{Conclusions}
\label{sec:conclusion}

We have investigated the role of interaction effects in the quantum geometry. By using a paradigmatic flat-band model, the fermionic Creutz ladder, it was found that the repulsive Hubbard interaction inhibits the quantum metric. Furthermore, by comparing the two extensions of the quantum metric to the interacting case available in the literature, namely the GQM and the DQM, with the MBQM obtained via exact diagonalization, we have found that the DQM provides a better fit to the exact diagonalization results. This conclusion applies both in the flat-band regime, which allows for an analytical treatment, and in the case of weakly dispersive bands, see Figs.~\ref{fig:g_comp_all_three_methods} and \ref{fig:disp-g}. Thus, our results suggest that the DQM offers a good approximation to the MBQM in cases where obtaining the exact many-body ground state is unfeasible.

We trace the difference between the GQM and the DQM to the different ways they account for the poles of the renormalized Green's functions. On a more fundamental level, there is a direct link between the DQM and the fidelity susceptibility, and consequently to the wave-function overlap like the MBQM. The GQM, while based on a natural extension from noninteracting systems, lacks the same intrinsic connection to the MBQM.

Physically, the interaction effects are manifested in the localization length, which being proportional to the quantum metric, vanishes at $U/t\to\infty$, see Fig.\ \ref{fig:strong_coupling} for the quantum metric at $U/t\gg1$. In addition, interaction-induced shifts of the original bands and replica bands are manifested in the spectral function and the optical conductivity.

Due to the considerable current interest in the quantum metric and its role in several physical phenomena, we expect our results to be important in situations where interaction corrections to the quantum metric itself are usually ignored. Among possible extensions of our work, we mention the calculation of the interacting quantum metric in different types of flat-band models as well as two- and three-dimensional systems. Of particular interest could be metallic systems and superconductors, where the quantum metric is manifested in the Drude weight and the superfluid weight. Prominent examples of such systems would be twisted bilayer graphene and semimetals with multifold band-crossings.

\begin{acknowledgments}
We acknowledge discussions with Riccardo Comin. This work was supported by the Research Council of Norway (RCN) through its Centres of Excellence funding scheme, Project No.\ 262633, ``QuSpin'', RCN Project No.\ 323766, as well as COST Action CA21144 ``Superconducting Nanodevices and Quantum Materials for Coherent Manipulation". KM was additionally supported by the DFG (SFB 1170) and the W{\"u}rzburg-Dresden Cluster of Excellence ct.qmat, EXC 2147 (Project-Id 390858490).
\end{acknowledgments}

\appendix

\begin{widetext}

\section{Calculation of self-energy}
\label{app:self-energy}

In this appendix, we provide the details of our calculations of the self-energies in the original and diagonal bases. We focus on the nontrivial contributions that cannot be represented as a shift of the Fermi energy.

\subsection{Original basis}
\label{app:self-energy-orig}

The interaction line corresponding to the Hubbard interaction in Eq.~\eqref{H_int_orig_local} and the Feynman diagrams corresponding to the nontrivial second-order contributions to the self-energy are given in Fig.~\ref{fig:self-energy-orig}. Other first- and second-order diagrams only shift the Fermi energy and, hence, can be neglected.

\begin{figure*}[b]
    \centering
    \includegraphics[height=0.15\textwidth]{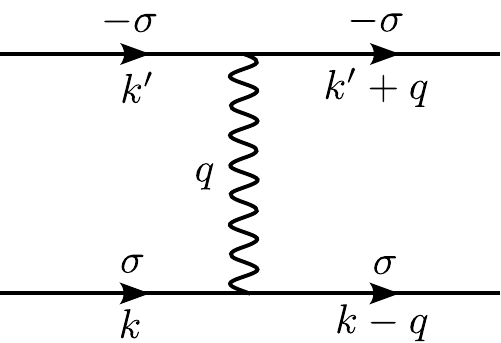}
    \includegraphics[height=0.15\textwidth]{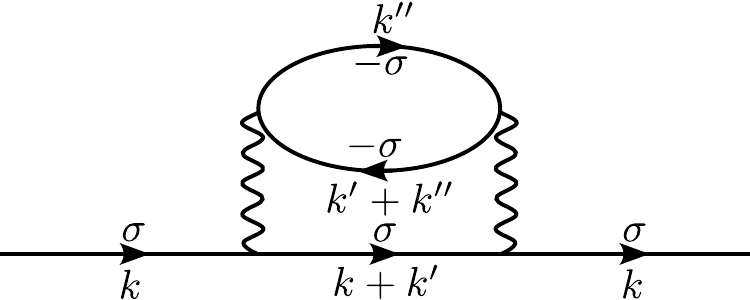}
    \includegraphics[height=0.15\textwidth]{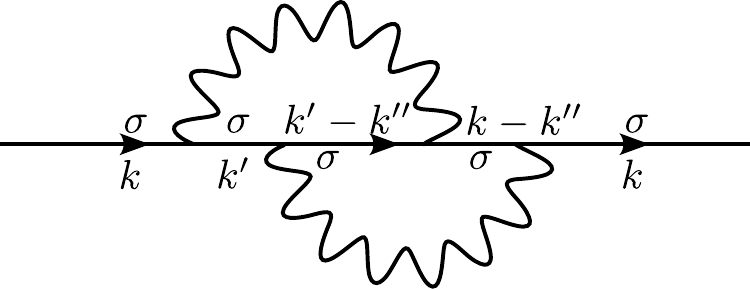}
    \caption{The interaction wavy line for the Hubbard interaction in Eq.~\eqref{H_int_orig_local} is shown in the left panel. Here, $\sigma$ denotes the spin projection. The direct (middle panel) and exchange (right panel) diagrams that provide nontrivial contributions to the self-energy $\Sigma^{\rm (orig)}(k,\omega)$ in the original basis.
    }
    \label{fig:self-energy-orig}
\end{figure*}

The Hubbard interaction between particles of opposite spins is $U_{\sigma_{1}\sigma_{2}; \sigma_{3} \sigma_{4}} =U \delta_{\sigma_{1}\sigma_{2}}\delta_{\sigma_{3}\sigma_{4}}(1-\delta_{\sigma_1 \sigma_3})$,
see the left panel in Fig.~\ref{fig:self-energy-orig}. Note that the corresponding interaction line does not flip the spin in the vertices. This property is implicitly used in the middle and right panels of Fig.~\ref{fig:self-energy-orig}.

The matrix element of the self-energy contribution from the direct diagram, see the middle panel in Fig.~\ref{fig:self-energy-orig}, is
\begin{eqnarray}
\label{self-energy-orig-Sigma}
&&\Sigma_{ab;\sigma_1\sigma_6}^{\rm (direct)}(k, i\omega_n) = - T^2\sum_{\left\{\sigma \right\}} \sum_{i\omega_{n'},i\omega_{n''}}  \int \frac{d k'}{2\pi} \int \frac{d k''}{2\pi} \delta_{\sigma_2\sigma_6} U_{\sigma_1 \sigma_2 \sigma_3 \sigma_4} G_{0, ab, \sigma_2}^{\rm (orig)}(k+k', i\omega_n +i\omega_{n'}) U_{\sigma_2 \sigma_5 \sigma_4 \sigma_3} G_{0, ab, \sigma_3}^{\rm (orig)}(k'', i\omega_{n''}) \nonumber\\
&&\times G_{0, ba, \sigma_4}^{\rm (orig)}(k'+k'', i\omega_{n'} +i\omega_{n''}) \nonumber\\
&&= -\sum_{\eta_1 =\pm} \sum_{\eta_2 =\pm} \sum_{\eta_3 =\pm} \sum_{\left\{\sigma \right\}} \int \frac{d k'}{2\pi} \int \frac{d k''}{2\pi} \delta_{\sigma_2\sigma_6} U_{\sigma_1 \sigma_2 \sigma_3 \sigma_4} U_{\sigma_2 \sigma_5 \sigma_4 \sigma_3} C_{\eta_1, ab, \sigma_2}(k+k') C_{\eta_2, ab, \sigma_3}(k'') C_{\eta_3, ba, \sigma_4}(k'+k'') \nonumber\\
&&\times F_{\eta_1,\eta_2,\eta_3}^{\rm (orig)}(k,k',k'')\nonumber\\
&&= -\delta_{\sigma_1\sigma_6} \sum_{\eta_1 =\pm} \sum_{\eta_2 =\pm} \sum_{\eta_3 =\pm} \int \frac{d k'}{2\pi} \int \frac{d k''}{2\pi} U_{\sigma_1 \sigma_1 -\sigma_1 -\sigma_1} U_{\sigma_1 \sigma_1 -\sigma_1 -\sigma_1} C_{\eta_1, ab, \sigma_1}(k+k') C_{\eta_2, ab, -\sigma_1}(k'') C_{\eta_3, ba, -\sigma_1}(k'+k'') \nonumber\\
&&\times F_{\eta_1,\eta_2,\eta_3}^{\rm (orig)}(k,k',k'')\nonumber\\
&&= -U^2 \delta_{\sigma_1\sigma_6} \sum_{\eta_1 =\pm} \sum_{\eta_2 =\pm} \sum_{\eta_3 =\pm} \int \frac{d k'}{2\pi} \int \frac{d k''}{2\pi} C_{\eta_1, ab, \sigma_1}(k+k') C_{\eta_2, ab, -\sigma_1}(k'') C_{\eta_3, ba, -\sigma_1}(k'+k'') F_{\eta_1,\eta_2,\eta_3}^{\rm (orig)}(k,k',k''),
\end{eqnarray}
where $\sum_{\left\{\sigma \right\}}$ is the summation over all internal spin indices $\sigma_i$ with $i=2,3,4,5$, $G_{0, \sigma}^{\rm (orig)}(k, i\omega_{n})$ is defined in Eq.~\eqref{creutz-G-orig-1}, $C_{\eta, \sigma}(k)$ is defined in Eq.~\eqref{creutz-G-orig-eta}, and
\begin{eqnarray}
\label{self-energy-orig-F}
&&F_{\eta_1,\eta_2,\eta_3}^{\rm (orig)}(k,k',k'') = T^2\sum_{i\omega_{n'}} \sum_{i\omega_{n''}} \frac{1}{i\omega_n+i\omega_{n'} -\epsilon_1}  \frac{1}{i\omega_{n''} -\epsilon_{2}} \frac{1}{i\omega_{n} +i\omega_{n''} -\epsilon_{3}} \nonumber\\
&&= T\sum_{i\omega_{n'}} \frac{1}{i\omega_n+i\omega_{n'} -\epsilon_1} \frac{n_{F}(\epsilon_2) -n_{F}(-i\omega_{n'} +\epsilon_{3})}{i\omega_{n'} +\epsilon_2 -\epsilon_3} = T\sum_{i\omega_{n'}} \frac{1}{i\omega_n+i\omega_{n'} -\epsilon_1}  \frac{n_{F}(\epsilon_2) +n_{B}(\epsilon_{3})}{i\omega_{n'} +\epsilon_2 -\epsilon_3} \nonumber\\
&&= -\frac{\left[n_{F}(\epsilon_1 -i\omega_n) -n_{F}(\epsilon_{3}-\epsilon_2)\right] \left[n_{F}(\epsilon_2) +n_{B}(\epsilon_{3})\right]}{i\omega_{n} -\epsilon_1 -\epsilon_2 +\epsilon_3} = \frac{\left[n_{B}(\epsilon_1) +n_{F}(\epsilon_{3}-\epsilon_2)\right] \left[n_{F}(\epsilon_2) +n_{B}(\epsilon_{3})\right]}{i\omega_{n} -\epsilon_1 -\epsilon_2 +\epsilon_3},
\end{eqnarray}
where $\epsilon_1 = \epsilon_{\eta_1, k+k', \sigma}$, $\epsilon_2 = \epsilon_{\eta_2, k'', \sigma}$, and $\epsilon_3 = \epsilon_{\eta_3, k'+k'', \sigma}$. The flat-band limit of Eq.~\eqref{self-energy-orig-F} is given in Eq.~\eqref{creutz-G-orig-F}.

Assuming the flat-band limit, Eq.~\eqref{self-energy-orig-Sigma} is rewritten as
\begin{eqnarray}
\label{self-energy-orig-Sigma-fb}
\Sigma_{ab;\sigma_1\sigma_2}^{\rm (direct)}(k,i\omega_n) &=& U^2\delta_{\sigma_1 \sigma_2} \sum_{\eta =\pm} \int \frac{d k'}{2\pi} \int \frac{d k''}{2\pi} C_{-\eta, ab, \sigma_1}(k+k') C_{-\eta, ab, -\sigma_1}(k'') C_{\eta, ba, -\sigma_1}(k'+k'') \frac{1}{i\omega_n +3\eta t}.
\end{eqnarray}
The integrals over momenta in Eq.~\eqref{self-energy-orig-Sigma-fb} are performed analytically. The final result is given in Eq.~\eqref{creutz-G-orig-Sigma}.

The contribution of the exchange diagram to the self-energy, see the right panel in Fig.~\ref{fig:self-energy-orig}, vanishes due to the spin structure of the Hubbard interaction. Indeed, the matrix elements of the corresponding contribution to self-energy $\Sigma^{\rm (exch)}(k, i\omega_n)$ is
\begin{eqnarray}
\label{analytics-original-KMP-Sigma-b-2-v1}
&&\Sigma_{ab; \sigma_1 \sigma_5}^{\rm (exch)}(k, i\omega_n) \nonumber\\
&&= T^2 \sum_{\left\{\sigma \right\}} \sum_{i\omega_{n'},i\omega_{n''}}  \int \frac{d k'}{2\pi} \int \frac{d k''}{2\pi} U_{\sigma_1 \sigma_2 \sigma_3 \sigma_4} U_{\sigma_1 \sigma_3 \sigma_4 \sigma_5} G_{0,ab, \sigma_1}^{\rm (orig)}(k', i\omega_{n'}) G_{0, ba, \sigma_3}^{\rm (orig)}(k'-k'', i\omega_{n'} -i\omega_{n''}) \nonumber\\
&&\times G_{0, ab, \sigma_4}^{\rm (orig)}(k-k'', i\omega_{n} -i\omega_{n''}) = \sum_{\left\{\sigma \right\}} \sum_{\eta_1 =\pm} \sum_{\eta_2 =\pm} \sum_{\eta_3 =\pm} \int \frac{d k'}{2\pi} \int \frac{d k''}{2\pi} U_{\sigma_1 \sigma_2 \sigma_3 \sigma_4} U_{\sigma_1 \sigma_3 \sigma_4 \sigma_5} C_{\eta_1, ab, \sigma_1}(k') C_{\eta_2, ba, \sigma_3}(k'-k'')  \nonumber\\
&&\times C_{\eta_3, ab, \sigma_4}(k -k'') F^{\rm (exch)}_{\eta_1,\eta_2,\eta_3}(k,k',k'') = 0.
\end{eqnarray}
%\end{widetext}
Here, $\sum_{\left\{\sigma \right\}}$ is the summation over all internal spin indices $\sigma_i$ with $i=2,3,4$ . In the summation, we used the structure of the spin-dependent interaction $U_{\sigma_1 \sigma_2 \sigma_3 \sigma_4} U_{\sigma_1 \sigma_3 \sigma_4 \sigma_5}=0$.

Therefore, the nontrivial contribution to the self-energy for the Hubbard interaction in the original basis is saturated by the direct type of the self-energy diagrams, see the middle panel in Fig.~\ref{fig:self-energy-orig} and Eq.~\eqref{self-energy-orig-Sigma}.

\subsection{Diagonal basis}
\label{app:self-energy-diag}

%\end{widetext}

In the diagonal basis, we perform a direct $S$-matrix expansion \cite{abrikosov, BruusFlensberg} starting from the Hubbard interaction in Eq.~\eqref{H_int_diag} to derive the self-energy. The only nontrivial contribution to the self-energy in the diagonal basis is the diagram shown in Fig.~\ref{fig:self-energy-diag}. The corresponding expression reads as
\begin{eqnarray}
\label{self-energy-diag-Sigma}
\Sigma_{\sigma}^{\eta\eta''}(k,i\omega_n) &=& -4 T^2 \int  \frac{dk'}{2\pi} \int\frac{dq}{2\pi} \sum_{i\omega_{n'},i\omega_\nu} \sum_{\eta_1 \eta_2 \eta'} U_{k+q, k'-q, k', k}^{\eta_1 \eta_2 \eta' \eta'',\sigma} U_{k', k, k+q, k'-q}^{\eta' \eta \eta_1 \eta_2,-\sigma}  G_{0, \eta',-\sigma}^{\rm (diag)}(k',i\omega_{n'}) G_{0, \eta_1,\sigma}^{\rm (diag)}(k+q,i\omega_{n}+i\omega_\nu)
\nonumber\\
&\times& G_{0,\eta_2,-\sigma}^{\rm (diag)}(k'-q,i\omega_{n'}-i\omega_\nu),
\end{eqnarray}
where the interaction strength $U_{k+q, k'-q, k', k}^{\eta_1 \eta_2 \eta' \eta,\sigma}$ is defined in Eq.~\eqref{H_int-U-def}.

%\begin{widetext}

\begin{figure}[b]
    \centering
    \includegraphics[height=0.125\textwidth]{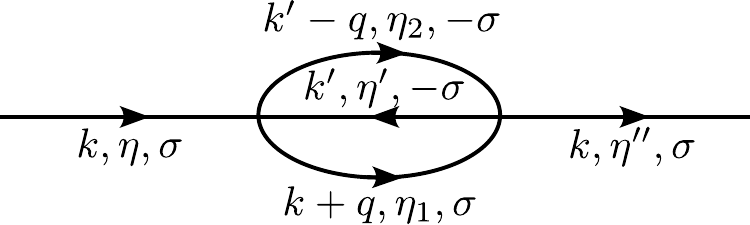}
    \caption{The second-order diagram describing the nontrivial contribution to the self-energy in the diagonal basis.}
    \label{fig:self-energy-diag}
\end{figure}

By using Eq.~\eqref{creutz-G-diag-eta}, we sum over the Matsubara frequencies as
\begin{eqnarray}
\label{analytics-original-CvG-Sigma-Matsubara}
&&F_{\eta',\eta_1,\eta_2}^{\rm (diag)}(k,k',q) = T^2 \sum_{i\omega_{n'}i\omega_\nu} \frac{1}{i\omega_{n'}-\epsilon_{1}}\frac{1}{i\omega_{\nu}+i\omega_{n}-\epsilon_{2}} \frac{1}{i\omega_{n'}-i\omega_{\nu}-\epsilon_{3}}
=  T\sum_{i\omega_{\nu}} \frac{-1}{i\omega_{\nu}+i\omega_{n}-\epsilon_{2}} \frac{n_F(\epsilon_1)-n_F(i\omega_\nu+\epsilon_3)}{i\omega_\nu -\epsilon_1+\epsilon_3} \nonumber\\
&&=  \frac{[n_F(\epsilon_1)+n_B(\epsilon_3)][n_F(\epsilon_2-i\omega_n)-n_F(\epsilon_1-\epsilon_3)]}{i\omega_n +\epsilon_1-\epsilon_2-\epsilon_3} = -\frac{[n_F(\epsilon_{1})+n_B(\epsilon_{3})][n_F(\epsilon_{1}-\epsilon_{3})+n_B(\epsilon_2)]}{i\omega_n +\epsilon_{1}-\epsilon_{2}-\epsilon_{3}},
\end{eqnarray}
where $\epsilon_1=\epsilon_{k'\eta',-\sigma}$, $\epsilon_2=\epsilon_{k+q,\eta_1,\sigma}$, and $\epsilon_3 = \epsilon_{k'-q,\eta_2,-\sigma}$. The last line is given as Eq.~\eqref{creutz-G-diag-F} in the main text. Performing the integrations over momenta and using Eq.~\eqref{analytics-original-CvG-Sigma-Matsubara}, the final result for the self-energy is given in Eq.~\eqref{creutz-G-diag-Sigma-def}.

%\begin{widetext}
\section{Quantum metric in flat-band limit}
\label{app:g_steps}

In this Appendix, we provide the details of our calculations of the quantum metric in the interacting flat-band Creutz ladder.

\subsection{Generalized quantum metric}
\label{app:g_steps-GQM}

Let us start with the generalized quantum metric (GQM) discussed in Sec.~\ref{sec:quantum_metric-GQM}. The quantum metric is related to the optical conductivity via Eq.~\eqref{metric_cond_relation}, see also Eq.~\eqref{metric_cond_relation-gk}. First, we insert the spectral functions in the original basis given in Eq.\ \eqref{creutz-G-orig-spectral} into the momentum integral of the trace in Eq.\ \eqref{Pi_def}. The integral is then carried out
\begin{align}
\label{g_steps-GQM-Pi}
\int\frac{\mathrm{d}k}{2\pi} \mathrm{tr}\big[J(k) A^{\rm (orig)}(k,\omega) J(k) A^{\rm (orig)}(k,\omega') \big] &= 4t^2\mathrm{sgn}\left(\Omega_U(\omega)\right)\mathrm{sgn}\left(\Omega_U(\omega')\right)
[\Omega_U(\omega)\Omega_U(\omega') - t_U(\omega)t_U(\omega')] \nonumber \\
&\times\df{\Omega^2_U(\omega)-t_U^2(\omega)}\delta(\Omega^2_U(\omega')-t_U^2(\omega')),
\end{align}
where $\Omega_U(\omega)$ and $t_{U}(\omega)$ are defined in Eqs.~\eqref{creutz-G-orig-OmegaU} and \eqref{creutz-G-orig-tU}, respectively; we performed analytical continuation $i\omega_n\to \omega +i0^{+}$. We use the standard relation to transform each of the delta functions in Eq.~\eqref{g_steps-GQM-Pi}:
\begin{equation}
\label{g_steps-GQM-df}
\df{\Omega^2_U(\omega)-t_U^2(\omega)}  = \sum_{\eta=\pm}\sum_{\alpha=1,2,3} \frac{\df{\omega -\eta \omega_{\alpha}}}{2\left|\Omega_U(\omega)\Omega'_U(\omega) -t_U(\omega)t_U'(\omega)\right|},
\end{equation}
where $\omega_{\alpha}$ with $\alpha=1,2,3$ are roots of the cubic equation $\Omega_U(\omega_{\alpha})=t_U(\omega_{\alpha})$; the full set of roots includes both $\omega_{\alpha}$ and $-\omega_{\alpha}$ which is accounted by $\sum_{\eta=\pm}$. The roots in the leading nontrivial order in $U/t$ are given in Eq.~\eqref{creutz-G-omega-roots}.

Assuming the zero-temperature limit, the following expression for the optical conductivity is derived
\begin{eqnarray}
\label{g_steps-GQM-sigma}
\RE{\sigma(\Omega)} &=& -\frac{4\pi t^2}{\Omega} \sum_{\alpha,\beta=1,2,3} \sum_{\eta_1,\eta_2=\pm}\int_{-\infty}^{\infty} d\omega \int_{-\infty}^{\infty} d\omega' \left[\theta{(-\omega)} -\theta{(-\omega')}\right] \delta(\omega -\omega' -\Omega) \left[\Omega_{U}(\omega)\Omega_{U}(\omega') -t_{U}(\omega) t_{U}(\omega')\right] \nonumber\\
&\times& \mathrm{sgn}(\Omega_{U}(\omega)) \mathrm{sgn}(\Omega_{U}(\omega')) \frac{\delta(\omega -\eta_1 \omega_{\alpha}) \delta(\omega' -\eta_2 \omega_{\beta})}{4\left|\Omega_{U}(\omega)\Omega_{U}'(\omega)
 -t_U(\omega)t_U'(\omega)\right| \left|\Omega_{U}(\omega')\Omega_{U}'(\omega') -t_{U}(\omega')t_{U}'(\omega') \right|} \nonumber\\
&=& -\frac{\pi t^2}{\Omega} \sum_{\alpha,\beta=1,2,3} \sum_{\eta_1,\eta_2=\pm} \left[\theta{(-\eta_1\omega_{\alpha})} -\theta{(-\eta_2\omega_{\beta})}\right]
\df{\eta_1\omega_{\alpha} -\eta_2\omega_{\beta} -\Omega} \left[\Omega_{U}(\eta_1\omega_{\alpha})\Omega_{U}(\eta_2\omega_{\beta}) -t_{U}(\eta_1\omega_{\alpha})t_{U}(\eta_2\omega_{\beta})\right] \nonumber\\
&\times& \frac{\sign{\Omega_{U}(\eta_1\omega_{\alpha} )} \sign{\Omega_{U}(\eta_2\omega_{\beta} )}}{\left|\Omega_{U}(\eta_1\omega_{\alpha} )\Omega_{U}'(\eta_1\omega_{\alpha}) -t_{U}(\eta_1\omega_{\alpha} )t_{U}'(\eta_1\omega_{\alpha})\right| \left|\Omega_{U}(\eta_2\omega_{\beta} )\Omega_{U}'(\eta_2\omega_{\beta}) -t_{U}(\eta_2\omega_{\beta} )t_{U}'(\eta_2\omega_{\beta})\right|}\nonumber\\
&=& -\frac{\pi t^2}{\Omega} \sum_{\alpha,\beta=1,2,3} \sum_{\eta_1,\eta_2=\pm} \left[\theta{(-\eta_1\omega_{\alpha})} -\theta{(-\eta_2\omega_{\beta})}\right]
\df{\eta_1\omega_{\alpha} -\eta_2\omega_{\beta} -\Omega} \frac{\left(\eta_1 \eta_2 -1\right) \eta_1 \eta_2}{\left|\Omega_{U}'(\omega_{\alpha}) -t_{U}'(\omega_{\alpha}) \right| \left|\Omega_{U}'(\omega_{\beta}) -t_{U}'(\omega_{\beta})\right|}\nonumber\\
&=& -\frac{2\pi t^2}{\Omega} \sum_{\alpha,\beta=1,2,3} \sum_{\eta_1=\pm} \left[\theta{(-\eta_1\omega_{\alpha})} -\theta{(\eta_1\omega_{\beta})}\right]
 \frac{\df{\eta_1(\omega_{\alpha} +\omega_{\beta}) -\Omega}}{\left|\Omega_{U}'(\omega_{\alpha}) -t_{U}'(\omega_{\alpha})\right| \left|\Omega_{U}'(\omega_{\beta}) -t_{U}'(\omega_{\beta})\right|}.
\end{eqnarray}

The GQM then follows from Eq.~\eqref{metric_cond_relation}:
\begin{eqnarray}
\label{g_steps-GQM-g}
g^{\rm GQM} &=& \int_{-\infty}^{\infty} d\Omega \frac{\theta{(\Omega)}}{\Omega} \frac{\sigma(\Omega)}{\pi} = -2t^2 \sum_{\alpha,\beta=1,2,3} \sum_{\eta_1=\pm} \frac{\theta{\left(\eta_1(\omega_{\alpha} +\omega_{\beta})\right)} \left[\theta{(-\eta_1\omega_{\alpha})} -\theta{(\eta_1\omega_{\beta})}\right]}{\left(\omega_{\alpha} +\omega_{\beta}\right)^2 \left|\Omega_{U}'(\omega_{\alpha}) -t_{U}'(\omega_{\alpha})\right| \left| \Omega_{U}'(\omega_{\beta}) -t_{U}'(\omega_{\beta})\right|}\nonumber\\
&=&-2t^2 \sum_{\alpha,\beta=1,2,3} \frac{\sign{\omega_{\alpha} +\omega_{\beta}} \left[\theta{(-\omega_{\alpha})} -\theta{(\omega_{\beta})}\right]}{(\omega_{\alpha}+\omega_{\beta})^2 \left|\Omega_{U}'(\omega_{\alpha}) -t_{U}'(\omega_{\alpha})\right| \left|\Omega_{U}'(\omega_{\beta}) -t_{U}'(\omega_{\beta})\right|}.
\end{eqnarray}
Using the symmetry with respect to $\alpha \leftrightarrow \beta$, the final result given in Eq.~\eqref{GQM-g-fin} is obtained.

\subsection{Dressed quantum metric}
\label{app:g_steps-DQM}

Let us now calculate the dressed quantum metric (DQM); see Sec.~\ref{sec:quantum_metric-DQM} for the discussion and final results. Assuming the zero-temperature limit in Eq.~\eqref{g_def_CvG} and using Eq.~\eqref{berry_curv_solved}, we obtain
\begin{eqnarray}
\label{g_steps-DQM-g}
g^{\rm DQM}(k) &=& \frac{1}{2} \sum_{\eta =\pm} \int_0^{\infty}d\Omega \int_{-\infty}^{\infty} d\omega A_{\eta}^{\rm (diag)}(k,\omega) A_{-\eta}^{\rm (diag)}(k,\omega+\Omega) \left[\theta{(\omega)} -\theta{(-\omega -\Omega)}\right] \nonumber\\
&=& \frac{1}{2} \sum_{\eta=\pm} \int_0^{\infty}d\Omega \int_{-\Omega}^{0} d\omega \df{\Omega_{U}(\omega) -\eta t_{U}(\omega)} \df{\Omega_{U}(\omega +\Omega) +\eta t_{U}(\omega+\Omega)} \nonumber\\
&=& \frac{1}{2} \sum_{\alpha,\beta=1,2,3} \sum_{\eta=\pm} \int_0^{\infty}d\Omega \int_{-\Omega}^{0} d\omega \frac{\df{\omega -\eta \omega_{\alpha}}}{\left|\Omega_{U}'(\omega) -\eta t_{U}'(\omega)\right|} \frac{\df{\omega+\Omega +\eta \omega_{\beta}}}{\left|\Omega_{U}'(\omega+\Omega) +\eta t_{U}'(\omega+\Omega)\right|}\nonumber\\
&=&\frac{1}{2} \sum_{\alpha,\beta=1,2,3} \sum_{\eta=\pm} \int_0^{\infty}d\Omega \frac{\theta{(-\eta \omega_{\alpha})} -\theta{(-\Omega -\eta \omega_{\alpha})}}{\left|\Omega_{U}'(\eta\omega_{\alpha}) -\eta t_{U}'(\eta\omega_{\alpha})\right| \left|\Omega_{U}'(\eta\omega_{\alpha}+\Omega) +\eta t_{U}'(\eta\omega_{\alpha}+\Omega)\right|} \df{\Omega +\eta (\omega_{\beta}+\omega_{\alpha})} \nonumber\\
&=& \frac{1}{2} \sum_{\alpha,\beta=1,2,3}\sum_{\eta=\pm} \theta{\left(-\eta (\omega_{\alpha}+\omega_{\beta})\right)} \frac{\theta{(-\eta \omega_{\alpha})} -\theta{(\eta \omega_{\beta})}}{\left|\Omega_{U}'(\eta\omega_{\alpha}) -\eta t_{U}'(\eta\omega_{\alpha})\right| \left|\Omega_{U}'(-\eta\omega_{\beta}) +\eta t_{U}'(-\eta\omega_{\beta})\right|}
\nonumber\\
&=& \frac{1}{2} \sum_{\alpha,\beta=1,2,3} \frac{ \theta{\left(\omega_{\alpha}+\omega_{\beta}\right)} \left[\theta{(\omega_{\alpha})} -\theta{(-\omega_{\beta})}\right]
+\theta{\left(-\omega_{\alpha}-\omega_{\beta}\right)} \left[\theta{(-\omega_{\alpha})} -\theta{(\omega_{\beta})}\right]}{\left|\Omega_{U}'(\omega_{\alpha}) - t_{U}'(\omega_{\alpha})\right| \left|\Omega_{U}'(\omega_{\beta}) - t_{U}'(\omega_{\beta})\right|}
\nonumber\\
&=& \frac{1}{2} \sum_{\alpha,\beta=1,2,3} \frac{\sign{\omega_{\alpha}+\omega_{\beta}} \left[\theta{(\omega_{\alpha})} -\theta{(-\omega_{\beta})}\right]}%{\left|\Omega_{U,+}'(\omega_{\alpha}) \Omega_{U,+}'(\omega_{\beta})\right|}
{\left|\Omega_{U}'(\omega_{\alpha}) - t_{U}'(\omega_{\alpha})\right| \left|\Omega_{U}'(\omega_{\beta}) - t_{U}'(\omega_{\beta})\right|},
\end{eqnarray}
where we used the flat-band limit in the second line.
Moreover, we used $\Omega_{U}'(-\eta \omega_{\alpha}) +\eta t_{U}'(-\eta \omega_{\alpha})=\Omega_{U}'(\eta\omega_{\alpha}) - \eta t_{U}'(\eta \omega_{\alpha})=\Omega_{U}'(\omega_{\alpha}) - t_{U}'(\omega_{\alpha})$ in the penultimate line.
Finally, using the symmetry with respect to $\alpha \leftrightarrow \beta$, one arrives at Eq.~\eqref{g_k_cvg_solved}.

\end{widetext}

\bibliography{main}

\end{document}